\documentclass[preprint,12pt]{elsarticle}

\usepackage{amssymb}

\newcounter{bla}

\journal{Computer Physics Communications}

\def\aplt{\ {\raise-.5ex\hbox{$\buildrel<\over\sim$}}\ }

\begin{document}

\begin{frontmatter}



\title{{\it DIAPHANE}: a portable radiation transport library for astrophysical applications}




\author[aff1,aff2]
{{Darren S.} { Reed}}
\ead{darren.reed@uzh.ch}

\author[aff3]{{Tim} {Dykes}}
\author[aff4]{{Rub\'en} {Cabez\'on}}
\author[aff5]{{Claudio} {Gheller}}
\author[aff1]{{Lucio} {Mayer}}

\address[aff1]{
 {Institute for Computational Science, University of Z\"{u}rich},  
 {Winterthurerstrasse 190},                      %
 {CH-8057}                             
 {Z\"{u}rich},                              
{Switzerland}                                     
}

\address[aff2]{
 {S3IT, University of Z\"{u}rich},  
 {Winterthurerstrasse 190},                      %
 {CH-8057}                             
 {Z\"{u}rich},                              
 {Switzerland}                                     
}

\address[aff3]{
 {School of Creative Technologies, University of Portsmouth},  
 {Eldon Building, Winston Churchill Avenue},                      %
 {PO1 2DJ}                             
 {Portsmouth},                              
 {United Kingdom}                                     
}

\address[aff4]{
 {Center for Scientific Computing, sciCORE, University of Basel}
 {Klingelbergstrasse 61},
 {CH-4056}
 {Basel},
 {Switzerland}
}

\address[aff5]{
 {Swiss National Supercomputing Centre, CSCS-ETHZ}
 {Via Trevano 131},
 {CH-6900}
 {Lugano},
 {Switzerland}
}


\begin{abstract}
One of the most computationally demanding aspects of the hydrodynamical modelling of Astrophysical phenomena is the transport of energy by radiation or relativistic particles.  Physical processes involving energy transport are ubiquitous and of capital importance in many scenarios ranging from planet formation to cosmic structure evolution, including explosive events like core collapse supernova or gamma-ray bursts. Moreover, the ability to model and hence understand these processes has often been limited by the approximations and incompleteness in the treatment of radiation and relativistic particles. The {\scshape diaphane} project has focused in developing a portable and scalable library that handles the transport of radiation and particles (in particular neutrinos) independently of the underlying hydrodynamic code.  
In this work, we present the computational framework and the functionalities of the first version of the {\scshape diaphane} library, which has been successfully ported to three different smoothed-particle hydrodynamic codes, {\scshape gadget2}, {\scshape gasoline} and {\scshape sphynx}.  We also present validation of different
modules solving the equations of radiation and neutrino transport using
different numerical schemes.
\end{abstract}



%

\end{frontmatter}

\section{Introduction}
\label{introduction}

The goal of the {\scshape diaphane} library is to enable advances in computational modelling of complex astrophysical processes in which the transfer of energy by radiation or neutrinos is important.  {\scshape diaphane} has been supported by the {\scshape pasc}\footnote{http://www.pasc-ch.org/} initiative, which aims to promote the development of software that can fully utilize new and emerging supercomputing hardware technology. Astrophysics simulations are currently able to capture many of the most relevant hydrodynamical and gravitational processes involved in the formation and evolution of stars, planets, and galaxies.  However, our advancement has been limited in a number of physical situations where energy transport by radiation or neutrinos is important. This project aims to improve the situation by providing a library of radiation and neutrino transport routines covering a range of physical regimes relevant for a wide range of astrophyics simulations, and also able to be used simultaneously within a single simulation.  We hope that the modular and extensible nature of this library will facilitate community contributions in the form of additional physics routines and other improvements after its initial public release.  

There are two primary reasons that including radiation and neutrino transport (RT/NT) in astrophysical simulations has been limited in scope.  1) The characteristic timescale for radiative and neutrino processes is very short compared to that of hydrodynamic processes. This is because the speed of light is much faster than the sound speed -- light travels $\sim 10^4$ times faster than sound in the molecular clouds where stars are forming. This means that simulation timesteps must be very short in order to accurately model the problem, increasing the global computation time, in some cases, to scales that are unaffordable nowadays.  And 2), RT/NT modelling is complex due to the many different processes involved at the atomic scale (e.g.\ emission, absorption, scattering).  This complexity is increased by the fact that a vast majority of astrophysical scenarios are intrinsically three-dimensional (3D), which translates into solving a full Boltzmann transport equation in 6D. This situation raises 3D hydrodynamical simulations with full RT/NT modelling to problems that lie well within the sustained Exascale.  

To skirt the computational complexity, RT/NT codes must implement approximate algorithms.  Typically, approximated RT/NT algorithms work well only for a specific range of conditions, e.g.\ optically thin diffuse gas with a single radiative source.  Nevertheless, most astrophysical phenomena include a large dynamic range and a corresponding large effective range in optical depth, so are not suited to be modeled by a single RT/NT approximate algorithm.  The {\scshape diaphane} project is unique at the moment in the international astrophysics community as it gathers different RT/NT approximations under a common portable framework, while so far, individual hydro codes have used only one or two radiative transfer solvers tailored to very specific problems, in very specific regimes, and adapted to very specific hydrodynamics implementation.

As a consequence, our understanding of the complex astrophysical Universe is hindered by the lack of radiation and neutrino transport (RT/NT) in our computational modeling of star and galaxy formation; planet formation; supernova explosions; and black hole formation.  We note that both the short timescales and the complex physics algorithms are very similar for RT and NT; for this reason it is effective and convenient to model the transport of radiation and neutrinos within a single framework implemented as a single library.

Radiation transport is important for the dynamics of many astrophysical processes.  Radiative feedback from stars and black holes plays a dominant role in regulating the rate of gas accretion into galaxies and star formation within galaxies.  The accumulated regulation of star formation determines the observable galaxy properties such as stellar luminosity, color, and age.  Because radiation can couple across a range of scales and conditions, multiple modes of RT modelling capabilities allow us to solve new classes of problems.  For example, a planet-forming disk around a protostar requires modelling the star light from the protostar as well as diffusion of that radiation outward through the planet forming disk. Evolving the disk temperature accurately, in this case, is needed to follow the fragmentation of the disk which ultimately determines the number and masses of planets that form.

Neutrino transport is of capital importance for understanding the explosion mechanism of one of the most powerful events in the Universe: core collapse supernovas. During the last phases of the lives of massive stars ($M>8-10M_{sun}$), a fast collapse of the core drives the formation of a neutron star via a strong deleptonization. As a result, an incredibly large quantity of neutrinos of all species is emitted. The interaction of these neutrinos, produced by the newly born neutron star, with the ensuing shockwave is critical to adequately understanding how the energy reservoir is tapped to energize and produce the explosions that we observe. Additionally, in the later phases of the explosion, as well as in different scenarios like neutron star mergers, the emission of highly neutronized matter by neutrino-driven winds is very important for characterizing the contribution of these scenarios to the r-process elements and understanding the origin of the nuclear abundances observed in the Galaxy . Therefore, the correct simulation of the neutrino-matter interactions in 3D geometrically distorted scenarios is the cornerstone of our understanding to such fundamental questions.

\section{The {\it DIAPHANE} library}
\label{diaphane}

\subsection{Overview of Library Modules for RT/NT physical processes}
\label{overviewmodules}

{\bf Common Utilities: }  This module handles many fundamental operations, ranging from defining the data structure to calculating quantities such as the opacity or the optical depth of gas given its physical conditions; these can be used by other library modules or called directly from a hydro simulation code as needed.

{\bf Flux-Limited Diffusion (FLD): } This is a relatively simple yet powerful method to approximate radiative diffusion.  For this reason, FLD served as our `prototype' module and was the first module to be ported to the library and validated for each of the three hydrodynamical codes included in the project. FLD is treated as thermal conduction as in \cite{ClearyMonaghan1999} \cite{Mayeretal2007}. Therein, photons diffuse in the direction of the local energy gradient.  
FLD is most accurate when gas is optically thick, typical of dense star-forming or planet-forming gas.  In this regime,  the FLD energy transport rate matches the diffusion limit, governed by the random walk of photon scattering. 
 See Sec.~\ref{fldmodule} for more details.

{\bf Starrad: } This newly developed ray casting method tracks the direct propagation of energy from a heating source, such as a star, to the surrounding medium, by casting rays onto a spherical angular grid and numerically solving the radiative transfer equation considering the absorption of the medium along each ray. 
{\scshape starrad} is a new type of algorithm in the context of particle based
codes. This technique of grid-based ray-tracing has been only used in grid-based codes to our knowledge (e.g. {\scshape flash}, {\scshape enzo}).
See Sec.~\ref{starradmodule} for more details.  

{\bf Advanced Spectral Leakage (ASL) \cite{perego16}:}  
This is a novel approximate neutrino treatment that computes the neutrino cooling rates by interpolating local production and diffusion rates (relevant in optically thin and thick regimes, respectively) separately for discretized values of the neutrino energy. Neutrino trapped components are also modeled, based on equilibrium and timescale arguments. The better accuracy achieved by the spectral treatment allows a more reliable computation of neutrino heating rates in optically thin conditions. See Sec.~\ref{aslmodule} for more details.

{\bf Isotropic Diffusion Source Approximation (IDSA) \cite{liebendoerfer09}:}
IDSA is a more sophisticated scheme in which the distribution function of the neutrinos is decomposed into two components: trapped and streaming. Their separate evolution equations are coupled by a source term that converts trapped particles into streaming particles. The efficiency of the scheme, when compared with a more detailed solution of the Bolztmann equation, results from the freedom to use different approximations for each particle component.

{\bf TRAPHIC (TRAnsport PHotons In Cones) \cite{PawlikSchaye2011}:} 
{\scshape traphic} is designed to model radiation from multiple sources efficiently.  An example problem is to model the heating and ionization of gas in the inter-galactic medium during the stage of rapidly increasing galaxy formation in the first billion years of the Universe.  The {\scshape traphic} module has been ported and used in post-processing mode between GADGET3 and GASOLINE in recent work on galaxy formation within our collaboration; see \cite{Tamburelloetal2017}.

\begin{figure}[h!]
\includegraphics[width=\columnwidth]{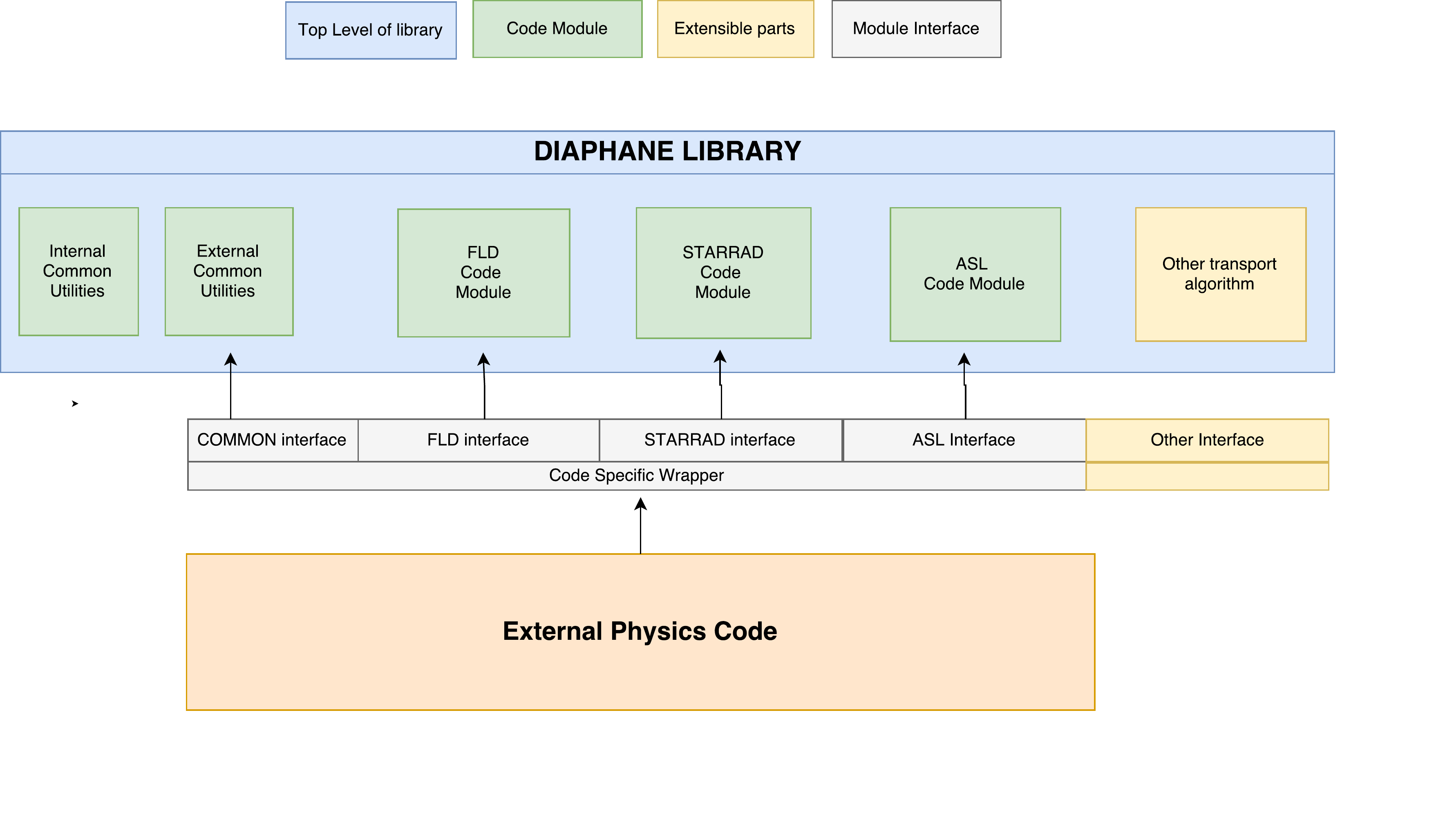}
  \caption{ Overall structure of the Diaphane Library.  Each module of the library contains an independent algorithm for radiation or neutrino transport.  The "Code Specific Wrapper" is the common API, and acts as a direct interface to a simulation code.  This wrapper, once called from a simulation code, then calls any module-specific interfaces (e.g., the FLD interface, the ASL interface, etc.), which contain any auxiliary functions needed to pass additional data to the modules that have been requested.  The special internal and external ``Common Utilties'' layers contain routines that are used by multiple modules, and are accessed from the code specific wrapper via the special "common interface". }
  \label{fig-structure}
 \end{figure}

\subsection{Implementation details}
\label{implementation}

In the following we outline some details of the {\scshape diaphane} library implementation.

\subsubsection{Library}
\label{lib}

The library is currently hosted in a Bitbucket repository. The attached link\footnote{https://bitbucket.org/diaphane/diaphane-library} points to the repository wiki, and will also contain the public release of {\scshape diaphane} in 2018 following a trial of validation and production runs conducted within the collaboration; v1.0 will consist of the first three modules (FLD, STARRAD, ASL) along with full documentation. The code will be published in the Computer Physics Communications (CPC) program library upon initial release.

The developmental focal points of the library are:

\begin{itemize}

\item{  \it{modular}}: Because there are a multitude of RT/NT algorithms and approximations, each having advantages or disadvantages in terms of accuracy and efficiency for specific physical conditions, it is convenient to design {\scshape diaphane} with a modular structure, so that adding new functionality is reasonably easy.

\item{ \it{extensible}}: Namely, we can add additional functions, support additional computational architectures, and couple with additional hydrodynamical codes -- all with minimal effort by exploiting the already existing framework.

\item{\it{maintainable}}: Well defined structure (Fig.~\ref{fig-structure}), naming conventions, focus on code clarity, GIT-based version control.

\item{ \it{portable}}: Multi-platform, support for multiple compilers, with a view toward planned accelerator support.

\item{\it{robust}}: Objective and efficient validation protocols.

\end{itemize}

\begin{figure}[h]
\includegraphics[width=.97\columnwidth]{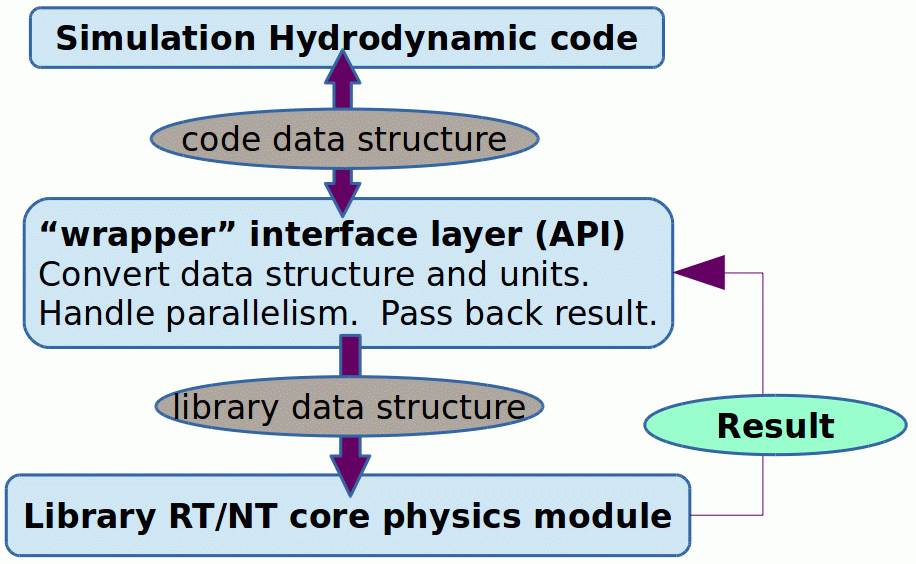}
  \caption{Functional schematic of the {\scshape diaphane} Library showing the relation between the library and the simulation code.  The wrapper layer contains the API to be called by a simulation code.  This layer translates the data structures of the main simulation code into those used by the library functions.   It has two levels.  The top-level code specific wrapper interface layer (API) is called from the simulation code; the top-level wrapper handles translations that are common to all modules, and it calls a second level module-specific interface for each requested module.  This module-specific interface layer contains auxiliary functions needed translate and pass any additional data that is needed by a specific module.  The core physics module layer contains all of the physics of the library and utilizes data structures that are standard to the library. }
  \label{fig-schematic}
 \end{figure}

The functional interactions between the library and a hydrodynamic code utilizing it are illustrated in Fig. \ref{fig-schematic}.
The core algorithmic layer of the library contains all the RT/NT physics modules, while the wrapper interface layer communicates with the main hydrodynamics code and with the module in the core physics layer.  When the interface layer is called by the hydrodynamics code, it then calls each RT/NT physics module that has been specified as part of the simulation.  The user will need to modify the wrapper layer to match the data structures and parallel methodology of their simulation code.  The top-level "code specific wrapper" acts as a commoon API that is called from a hydrodynamics code; it will call a code and module-specific interface to any module to be used.  This layer will contain any auxiliary functions needed by the specific module to interface with the hydrodynamics code.  We provide examples of this layer for a number of simulation codes, currently {\scshape gagdet2} \cite{springel2005}, {\small ChaNGa}/{\scshape gasoline} \cite{jetley2008,Wadsleyetal2004}, and {\scshape sphynx} \cite{cabezon17}, all three being smoothed particle hydrodynamics (SPH) codes.  Our goal is to make it as simple as practical to port the library to a new simulation code.  For routines that only require neighbor information, this is mainly a matter of passing a neighbor list and converting units.  However, for other algorithms that require information across domain boundaries, the module-specific interface layer must be made aware of how to access information on other domains.  Implementing this layer will require a reasonable working knowledge of the hydrodynamic code, but the examples already in the library will assist as a recipe for this process.  Once the interface layer has been written for a hydrodynamic code to use a particular module, the process of building a new library module from an existing physics routine should be a relatively simple process, which we can provide assistance with.  We hope that, over time, the community will add RT/NT modules to the library.  We also note that although we focus on utilizing the library during live simulations, it can also be used for pure post-processing calculations, if desired.

In our effort to ensure portability we have tested that the library, implemented in C/C++, can interface effectively with both {\scshape fortran} and C based codes.  {\scshape sphynx} is a {\scshape fortran} code while {\scshape gadget} and {\scshape gasoline} are C codes. We also plan to soon support {\small ChaNGa}, a closely related offshoot of {\scshape gasoline}, written mainly in C++. 
For every interface and shared data structure or procedure, we use the {iso\_c\_binding} module to ensure compiler-independent functionality.  This standard {\scshape c-fortran} interface is necessary because, among other reasons, different compilers will align data structures in memory differently.

In the end, calls to any library module are made via a common API defined in the interface layer, which then triggers a call to each RT/NT module that has been specified.  We can also provide a template upon request, to use the library coupled with a specific code, so that the required changes to the underlying hydrodynamical code are minimal.  Nevertheless, in some cases the {\scshape diaphane} library may require the hydrodynamical code to calculate and carry some additional data that is not usually part of the main calculation.  For example, an independent adaptive timestep length for radiation, discussed in the following paragraph, or the energy gradient, needed by FLD, is convenient to store as an additional per particle variable.

An important aspect regarding efficiency is that we also implemented the possibility of allowing {\scshape diaphane} to have an independent timestep, since the radiation timescales are usually much shorter than the hydrodynamic timescale. This allows the radiation calculations to be decoupled from the hydrodynamical calculations. One benefit is that the hydrodynamic calculations can be done at a much lower frequency if the simulated scenario and conditions allow it. Another benefit is that the radiation calculations have the possibility to be carried out on separate hardware, opening the opportunity to exploit accelerators exclusively for the much more costly RT/NT calculations.

The first prototype module (FLD) of the library utilizes 
a relatively brute-force approach, looping through each hydrodynamic element in the wrapper layer and applying the library algorithms one fluid element at a time.  This simple strategy means that only a small amount of information must be passed to the core library module, namely the properties of that element and its neighbors.
This means that the first library version had no integrated parallelism; communication was instead handled by the parallel functionality of the main simulation code. This strategy was sufficient for the relatively simple FLD module.  However, to implement the more complex {\scshape starrad} algorithm, it was necessary to provide means of accessing information across computational domains during the propagation of rays across the simulation volume. 
To accomplish this, we have utilized an 
abstraction layer for parallelism called the Machine Dependent Layer (MDL), which  is inherited from one of the partipating codes {\scshape gasoline}. The library in its final form will rely on MDL for communication, the portability and flexibility of which will be crucial for its success.
This paper focuses on the implementation of the modules and the general structure of the library while the parallel environment will be described in detail in a paper that will accompany the first public release of the code.

\subsection{Algorithmic descriptions}

We next describe the implementation of the three highlighted RT/NT modules that will be part of the first public release, with a focus on the relevant physics of the algorithms. 
This includes FLD as well as the two most recently incorporated modules, {\scshape starrad} and ASL, which are currently undergoing testing and validation.
The remaining modules of those that we have already mentioned in \S{\ref{overviewmodules}}
are at earlier stages in development, and will be added later.

\subsubsection{Flux-Limited Diffusion Module}
\label{fldmodule}

FLD models radiation transport through gas using a photon diffusion approximation \cite{Bodenheimeretal1990}.  It has been adapted to the SPH method following \cite{ClearyMonaghan1999, Mayeretal2007}.  The FLD energy rate of particle $a$ is:
\begin{equation}
\label{eqn:udot_sph}
  \dot{u}_a = \sum_b \frac{4 m_b}{\rho_a \rho_b} \frac{k_a k_b}{k_a + k_b} \left( T_a - T_b \right) \frac{\mathbf{r}_{ab} \cdot \nabla W}{|\mathbf{r}_{ab}|^2}
  \end{equation}
summed over all neighbors $b$ of mass $m_b$, separated by the vector $\mathbf{r}_{ab}$; $T_a$ and $\rho_a$ are the temperature and density, respectively. FLD energy flows along the local energy gradient, determined from the kernel gradient $\nabla W$. 
The thermal conductivity term,
\begin{equation}
  \label{eqn:k}
  k_a = \frac{16 \sigma}{\rho_a \kappa_a} \lambda_a T_a^3 ,
  \end{equation}
determines the effective diffusion coefficient; $\sigma$ is the Stefan-Boltzmann constant.  The flux limiter ($\lambda$) sets the rate of energy flux in the energy equation.  For optically thick gas, $\lambda$ matches that for the diffusion limit; for optically thin gas, $\lambda$ limits energy flux to the speed of light.  Opacity, $\kappa_a(\rho, T)$, is interpolated from a table of Rosseland mean opacities.  Finally, an FLD timestep constraint can be imposed such that
\begin{equation}
\label{eqn:taufld}
dt_{FLD} \le \eta_{FLD} \frac{u_a}{\dot{u}_{aFLD}},
\end{equation}
and similar timestepping can be imposed for the other radiative modules.

\subsubsection{STARRAD Module}
\label{starradmodule}

The {\scshape starrad} module is based upon a ray casting technique to describe stellar irradiation of the surrounding environment by solving the radiative transfer equation currently considering only absorption: 

\begin{equation}
\label{eqn:radiativetransport}
I_{x1} = I_{x0}\exp(-\tau(x0,x1)) 
\end{equation}

I.e. intensity at position \small{X1} is equal to the intensity at position \small{X0} reduced by the exponent of the optical depth between positions X0 and X1, Where optical depth $\tau$ is:

\begin{equation}
\label{eqn:tau}
\tau(x0,x1) = \int\limits_{x0}^{x1} \alpha(x)\ dx \
\end{equation}
 bit
and $\alpha$ (absorption coefficient) consists of the following, where $\kappa$ is opacity, and $\rho$ is the density at position \small{x}:
 
\begin{equation}
\label{eqn:alpha}
\alpha = \kappa \rho(x)
\end{equation}

\begin{figure}
\begin{center}
\includegraphics[width=\columnwidth]{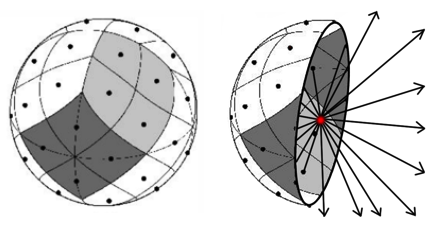} 
  \caption{HEALPix decomposition of sphere around heating source }
  \label{fig-healpix}

\includegraphics[width=.98\columnwidth]{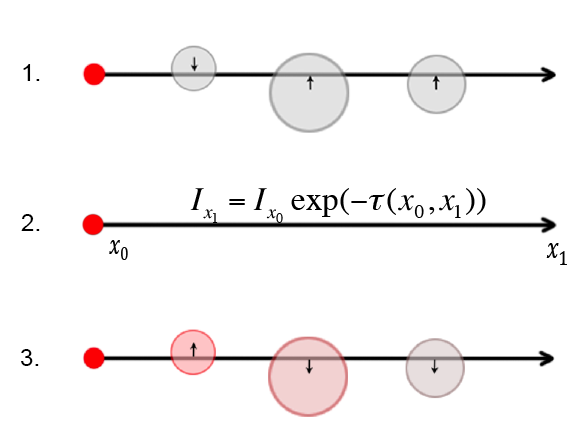} 
	\caption{Three phases of {\scshape starrad} algorithm; optical depth, ray integration, and heating, respectively. $\tau$ in the exponential refers to optical depth between points $X_{0}$ \& $X_{1}$}
  \label{fig-starrad}
\end{center}
\end{figure}

In order to calculate the direct radiative effect exerted by a heating source on its surroundings, a sphere of influence around the source is decomposed into pixels via HEALPix  \cite{Gorskietal2005} subdivision (Fig.~\ref{fig-healpix}), and a ray allocated per pixel.  Three tunable parameters  control the resolution of spherical grid: the density of the ray distribution (synonymous with the resolution of the HEALPix distribution), the total length of each ray, and the length of each ray segment.

The algorithm is then structured in three key phases (Fig.~\ref{fig-starrad}):

\begin{enumerate}
\item
{
\textbf{Optical depth}:
Sample the volume for optical depth contributions at discrete points along each ray, stored in ray segments. For SPH this is accomplished by iterating once through the particles, calculating particle-ray intersections and spreading the particle optical depth across the affected ray segments.
}
\item
{
\textbf{Ray integration}:
Integrate optical depth contributions and solve the radiative transport equation along each ray, resulting in a radiation intensity field surrounding the source.
}
\item
{
\textbf{Heating}:
Apply the radiation intensity field to heat the medium by the appropriate amount of energy absorption, as determined by the  optical depth calculation.
}
\end{enumerate}

The {\scshape starrad} module is currently being tested within the {\scshape gasoline} code, and so is parallelized through use of the Machine Dependent Layer (MDL) library, which serves as an abstraction layer for parallelism and is used by {\scshape gasoline}).  This approach is in line with the future direction of the {\scshape diaphane} library (See Sect.~\ref{conclusions}).

The parallel {\scshape starrad} model evenly distributes ray groups across processing elements (PEs). Ray segments are fetched on demand from remote PEs to write optical depth contributions, and read the intensity field for medium heating.  {\scshape starrad} makes use of fast caching mechanisms available in MDL to mitigate communication overhead. The HEALPix decomposition allows rays to be simply indexed to maintain spatial coherency when distributed across processors. Rays are integrated locally on each PE to convert the accumulated optical depth into a radiative intensity field.

Referring to the planet forming disk example of Sect.\ref{introduction}, in order to self-consistently model the evolution of gas surrounding a protostar, {\scshape starrad} can be used in a hybrid manner, similar to Klassen et al \cite{Klassenetal2014}, where each time-step exploits multiple methods of calculating radiative transfer.
Flux-limited diffusion, because it is well-suited for the optically thick environments of forming stellar systems, is the first module that we are coupling with {\scshape starrad}.  This enables us to split the radiation field into direct and indirect components, allowing higher accuracy in the evolution of the gas properties \cite{Murrayetal1994}. In this case, we model direct heating of the gas by the protostar's light with {\scshape starrad} and subsequently use the FLD module to follow diffusion of that heat in the optically thick disk, allowing the temperatures throughout the disk to be modeled much more realistically.

\subsubsection{ASL Module}
\label{aslmodule}
The Advanced Spectral Leakage (ASL) method \cite{perego16} is an approximate treatment for neutrinos and anti-neutrinos of all flavors, where we conjugate the usual positive aspects associated with leakage schemes (mainly, the reduced computational cost and flexibility), with an improved accuracy, obtained using a spectral approach (i.e., solving different leakage schemes, for different energy bins), modeling a neutrino trapped component in the optically thick region, and including a consistent absorption term obtained from the spectral cooling rates. ASL works in the following way:
\begin{enumerate}
\item compute the corresponding spectral emissivity, absorptivity or scattering rate at each fluid element position. The total emissivity and absorptivity are computed as the sum over all considered neutrino processes.
\item compute the local total mean free path of the neutrinos $\lambda_{\nu,tot}$ and the energy mean free path $\lambda_{\nu,en}$ as a geometrical average between the mean free path due to all absorption processes and that due only to highly inelastic processes (i.e. processes where the energies of the incoming and outgoing neutrinos is expected to differ considerably).
\item perform path integrals of both mean free paths to obtain the corresponding local optical depths $\tau_{\nu,tot}$ and $\tau_{\nu,en}$. According to their values we distinguish four different regimes: equilibrium-diffusive ($\tau_{\nu,tot}\gg 1$ and $\tau_{\nu,en}\gtrsim 1$), diffusive ($\tau_{\nu,tot}\gg 1$ but $\tau_{\nu,en}\lesssim 1$), semi-transparent ($\tau_{\nu,tot}\sim 1$), and free-streaming ($\tau_{\nu,tot}\lesssim 1$).
\item compute the trapped component variation rates of neutrino abundances and energies from the reconstructed neutrino distribution function adequately corrected by a cutoff at regimes other than equilibrium-diffusive.
\item evaluate the change of energy due to neutrino cooling and absorption, including and effective method to evaluate heating rates in optically thin regions based on the local spectral neutrino density.
\end{enumerate}

In the end, ASL provides rates of change of specific energy, neutrino abundances and energies, and a neutrino pressure that can be added to the momentum equation in order to take into account momentum transfer from the neutrino field to the matter field.

ASL has been implemented in the 1D code {\scshape agile} \cite{liebendorfer04}, and the 3D codes {\scshape fish} \cite{kaeppeli11} (Cartesian mesh) and {\scshape sphynx} (Cartesian SPH). It is worth to note that the implementation of the (more detailed) IDSA neutrino treatment will easily follow from the current ASL implementation. The reason for this is that IDSA follows the same structure presented above, but the last two steps. Instead of steps 4. and 5., IDSA constructs a streaming particle background field from all sources of the previous time step/iteration and solves a diffusion equation that locally takes these streaming neutrinos into account to tailor the interpolation between the optically thick and thin regimes.  Both methods, ASL and IDSA, have been extensively validated against {\scshape bolztran} \cite{mezzacappa93a,mezzacappa93b}, that solves the Boltzmann transport equation in 1D without assuming any approximation \cite{perego16,liebendoerfer09}.

We are currently working in the development of the wrapper layer for this module, to be followed by a series of tests that can be compared with previous calculations performed in the core collapse scenario.

\section{Validation}
\label{results}
\begin{figure}[h!]
\begin{tabular}{cc}
\includegraphics[width=.46\columnwidth]{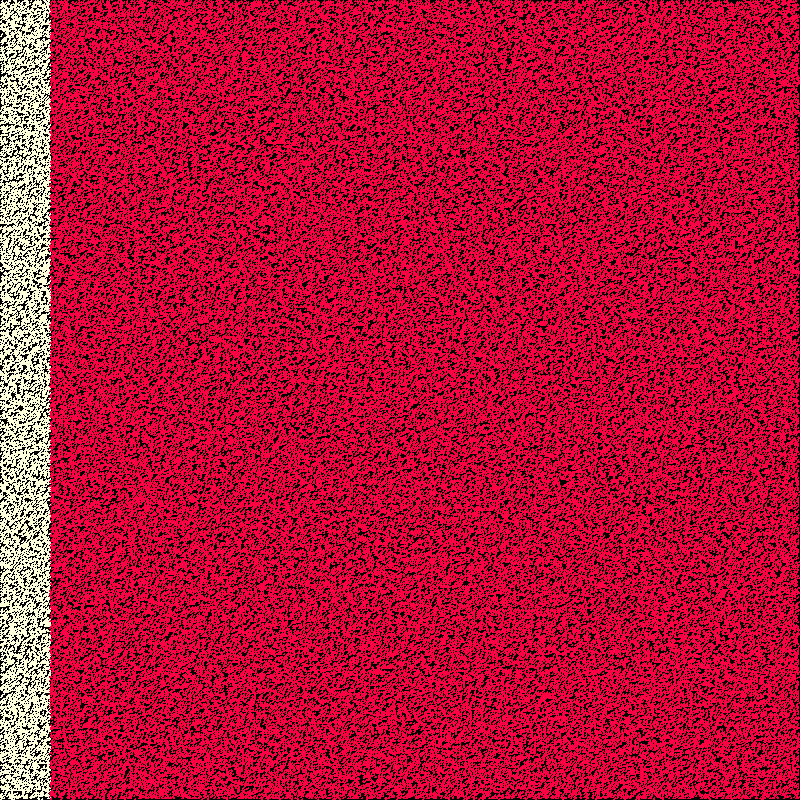} &
\includegraphics[width=.46\columnwidth]{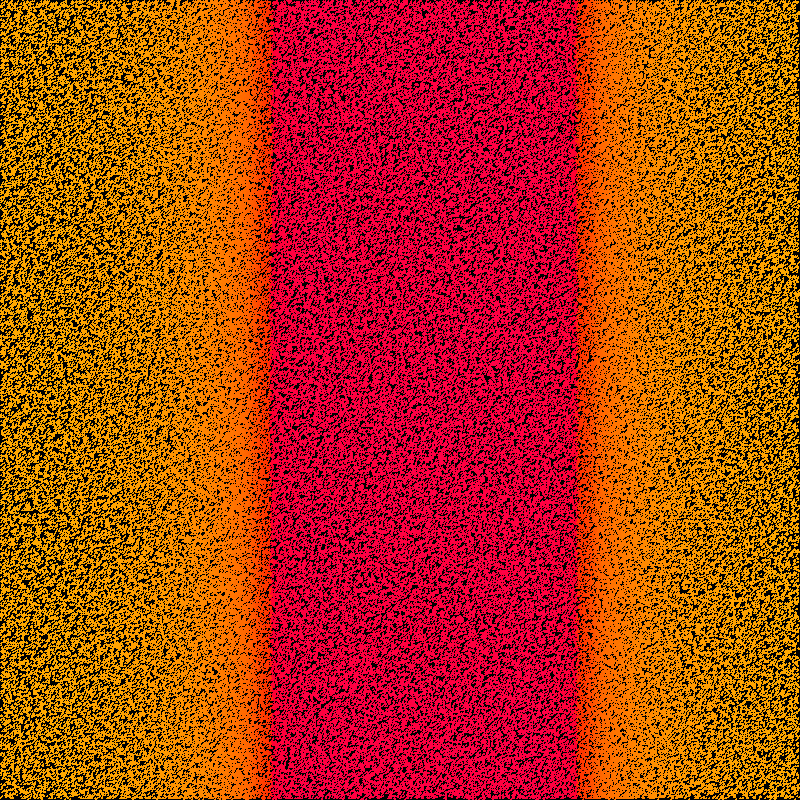} \\
\includegraphics[width=.46\columnwidth]{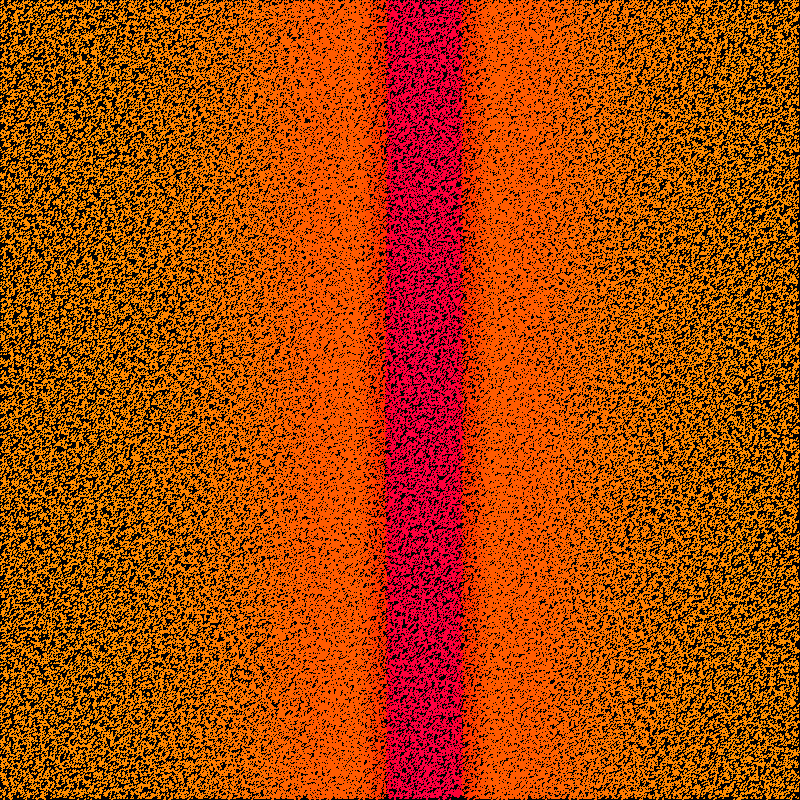} &
\includegraphics[width=.46\columnwidth]{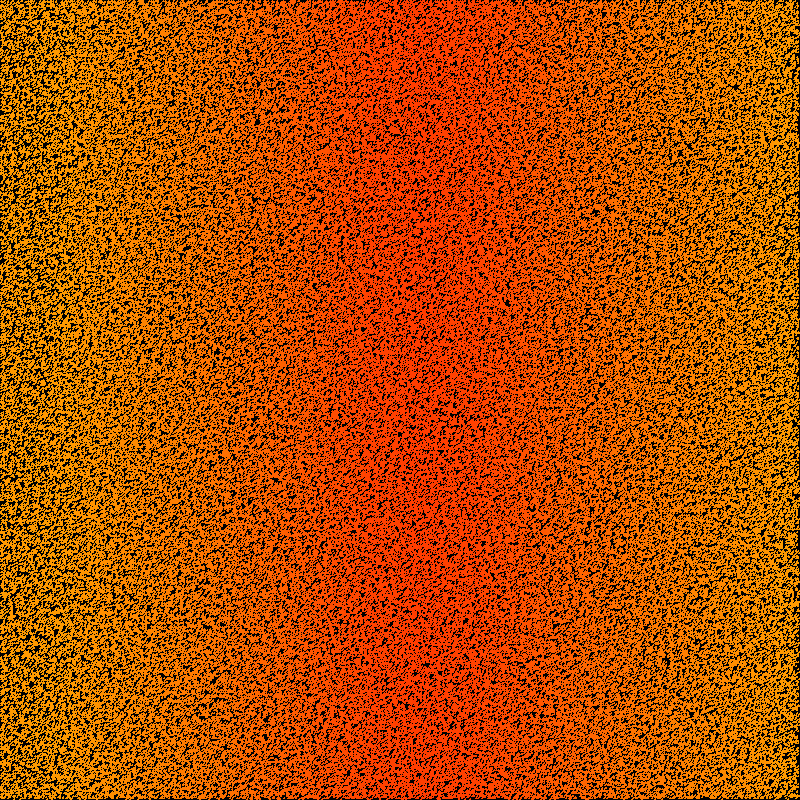} \\
\end{tabular}

  \caption{Image of cubical validation simulation for Flux-Limited Diffusion, colored by gas temperature.  Initial conditions are a 100K periodic cube of cosmic gas with a 1000K layer.  Time increases left to right and top to bottom; starting from upper left, the times shown are $t=0$, $50$, $100$, and $300$ ${\rm \times 10^{8}s}$). }
  \label{fig-cubepic}
 \end{figure}

\begin{figure}
\includegraphics[width=.8\columnwidth]{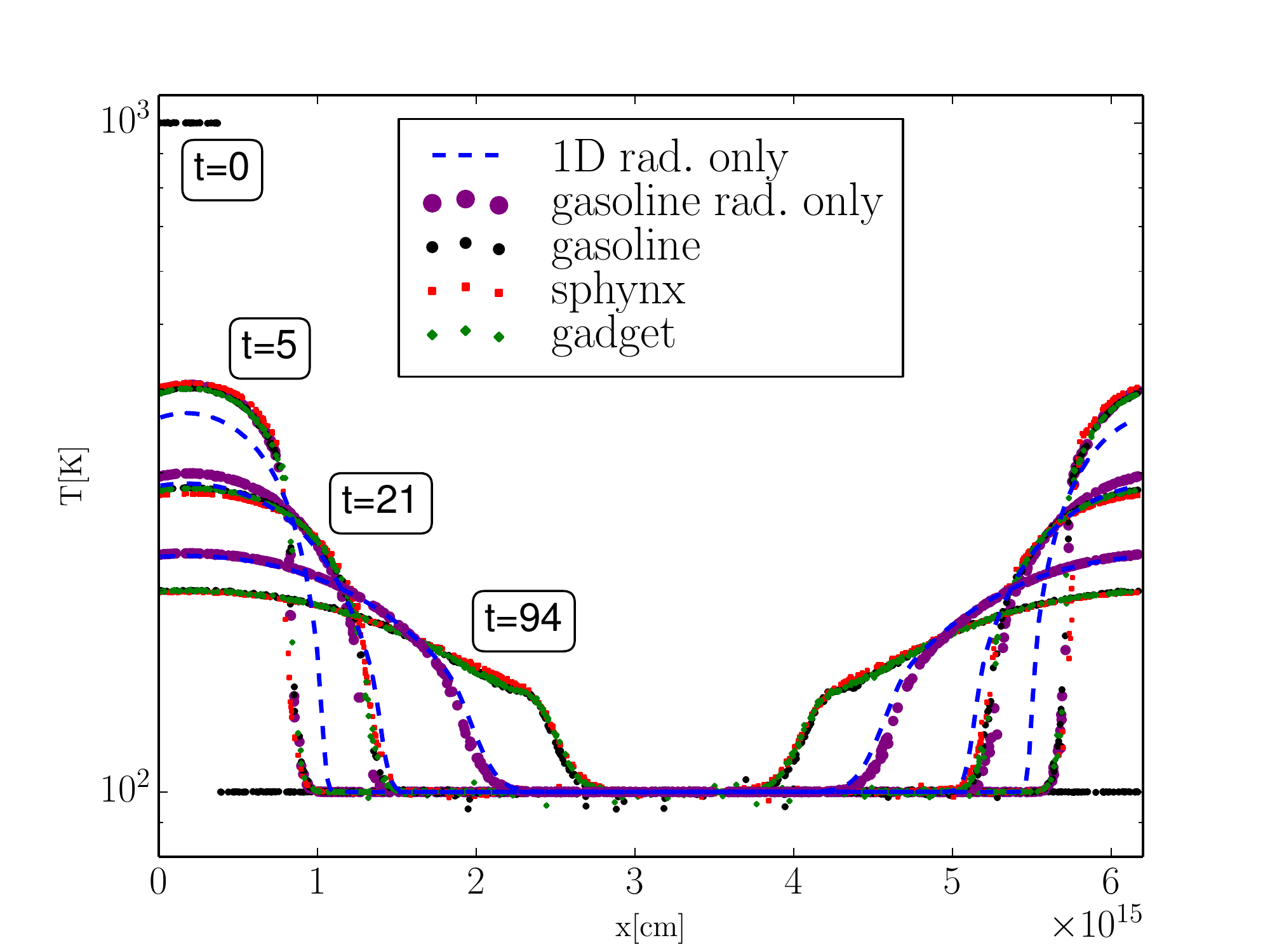}
  \caption{Cube test for Flux-Limited Diffusion.  Initial conditions are a 100K periodic cube of cosmic gas with a 1000K layer.  Plotted is a random 1 in 512 sampling of particles from each simulation.  The three hydrodynamic codes agree with each other.  The 1D semi-analytic solution does not include hydrodynamic forces.  The 3D run radiation only case (i.e. static particle positions) agrees with the 1D case.  The difference between the radiation-only simulation and the full hydrodynamic simulations at late times is due to thermal pressure.  Time plotted is in units of $10^{8}$s. 
  }
  \label{fig-cubetemp}
\includegraphics[width=.8\columnwidth]{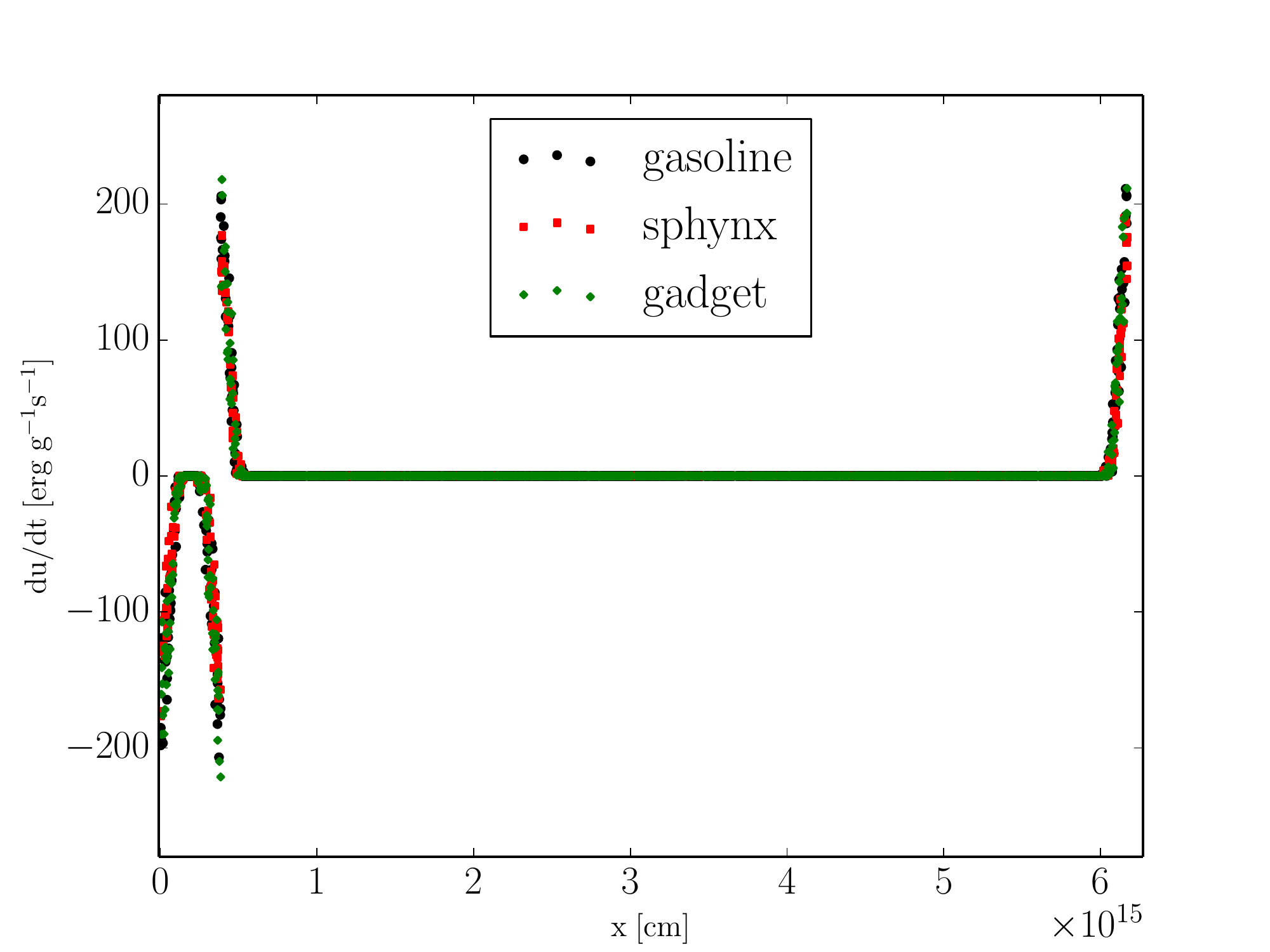}
  \caption{Cube test for Flux-Limited Diffusion heating/cooling rates verifying the independence of FLD rates with simulation code.  Particles plotted are a random 1 in 128 sampling of the simulation.  Shown at time $t=0$, energy flows across the temperature discontinuity.  }

  \label{fig-cubedudt}
 \end{figure}

We now present results of our code validation tests.
For each library module, we run a simple test problem using at least three different hydrodynamic codes.  For FLD, diffusion through a periodic cube of gas provides a useful test for conditions relevant for planet forming disks around stars.  The cube has uniform density ${\rm 5 \times 10^{-13}~g~cm^{-3}}$ and length ${\rm 6 \times 10^{15}~cm}.$ 
In Fig.~\ref{fig-cubepic} we show a projection of the particle distribution at four different times, from top-left to bottom-right respectively. Temperature is color-coded. As expected, the system evolves towards equilibrium, trying to homogenize the temperature profile via energy diffusion.  A more detailed plot of the evolution of the temperature profile is shown in Fig.~\ref{fig-cubetemp}. There, we plot the results obtained by coupling {\scshape diaphane} with three different SPH codes ({\scshape gadget}, {\scshape gasoline}, and {\scshape sphynx}) at different stages of the evolution.  We include a 1D semi-analytic simulation of diffusion as a verification of the FLD implementation.  Aside from an initial offset during the early relaxation phase of the simulation, the 1D solution agrees well with the static 3D solution (run with {\scshape gasoline} with particle positions fixed in space).  The effect of thermal pressure can be seen as a divergence between the static solutions and those that include acceleration due to hydrodynamic forces.
Fig.~\ref{fig-cubedudt} confirms that the rates of energy change of simulation particles are in good agreement among different simulation codes.

\begin{figure}[h!]
\begin{tabular}{cc}
\includegraphics[width=.46\columnwidth]{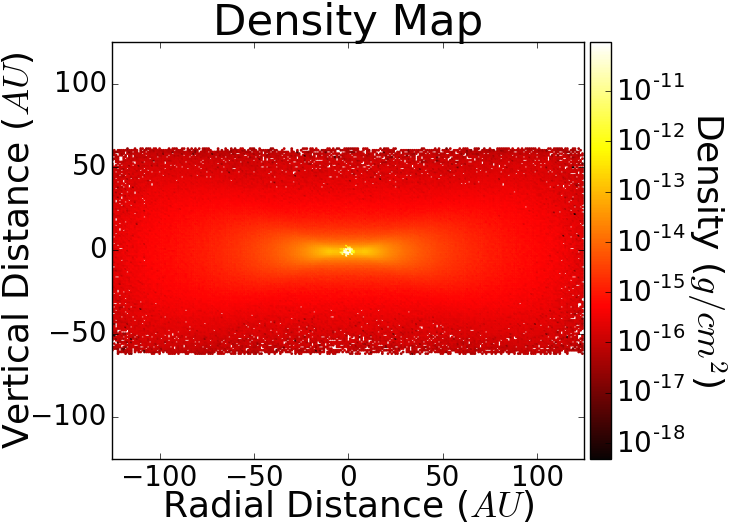} & 
\includegraphics[width=.46\columnwidth]{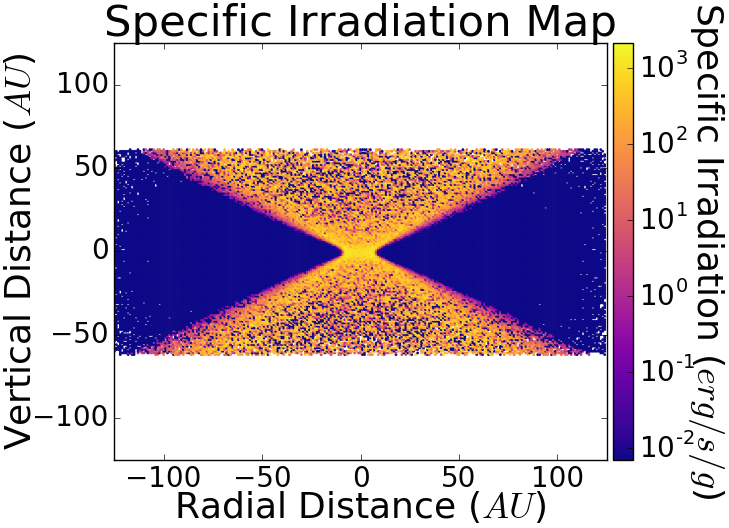} \\
\includegraphics[width=.46\columnwidth]{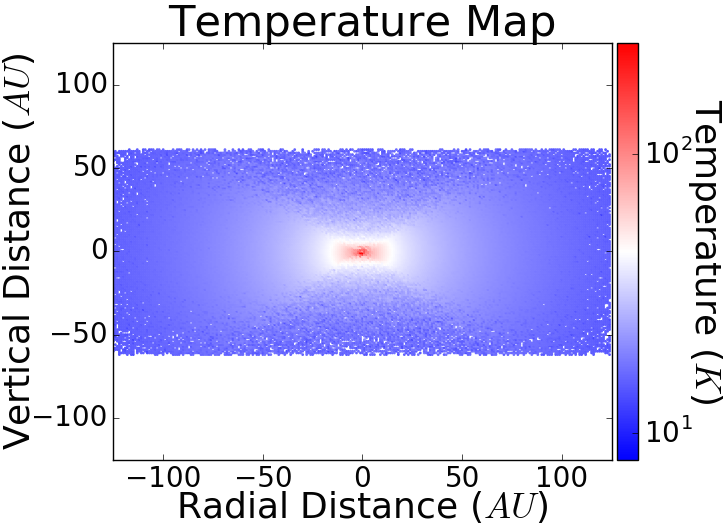} & 
\includegraphics[width=.46\columnwidth]{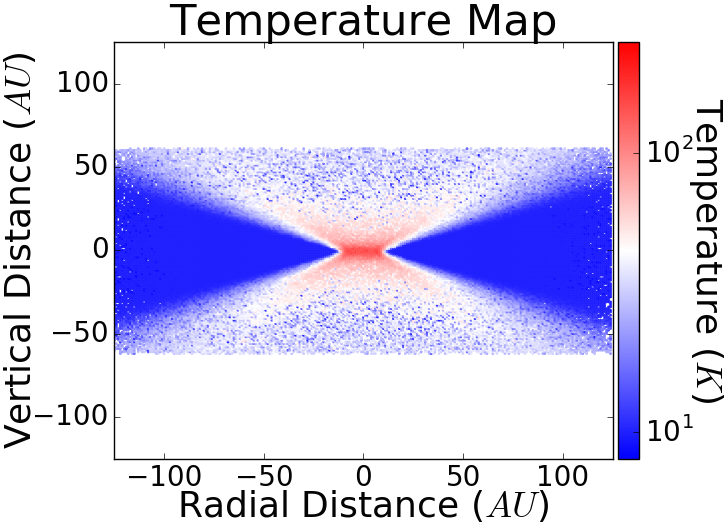} \\
\end{tabular}{}

 \caption{Color coded maps representing a {\scshape starrad} static irradiated disk test; initial density map (left, top), initial temperature map (left, bottom). Specific irradiation (energy per unit time per unit mass) during the first step (right, top), and temperature after 300 years evolution (right, bottom).}
  \label{fig-starrad-profiles}
 \end{figure}

For {\scshape starrad}, we run a validation test consisting of a sun-like star enveloped in a gaseous disk. The disk is evolved statically, i.e. without hydrodynamical accelerations, from an initial disk shown in Fig.~\ref{fig-starrad-profiles} (top/bottom left). Initial conditions are generated using an iterative procedure that enforces initial hydro-static equilibrium by taking into account all forces (pressure, gravitational and centrifugal, see Rogers \& Wadsley 2011 \cite{rogerswadsley2011}). 

As {\scshape starrad} computes only heating, some form of cooling is required in order to evolve to thermal equilibrium. Whilst the FLD and {\scshape starrad} coupling process (see Sect~\ref{starradmodule}) is on-going, we have paired {\scshape starrad} with a standard cooling module within {\scshape gasoline} \cite{Galvagnietal2012} \cite{szulagyietal2017}. This cooling model adopts a local optical depth approach which recovers the optically thick and optically thin limits of the cooling rate (essentially equivalent to the diffusion limit employed by FLD for large optical depths). This takes advantage of the optical depth calculation performed by {\scshape starrad}, which will be a common module of the library also usable by other library modules.

\begin{figure}[h!]
\begin{center}  
\includegraphics[width=.97\columnwidth]{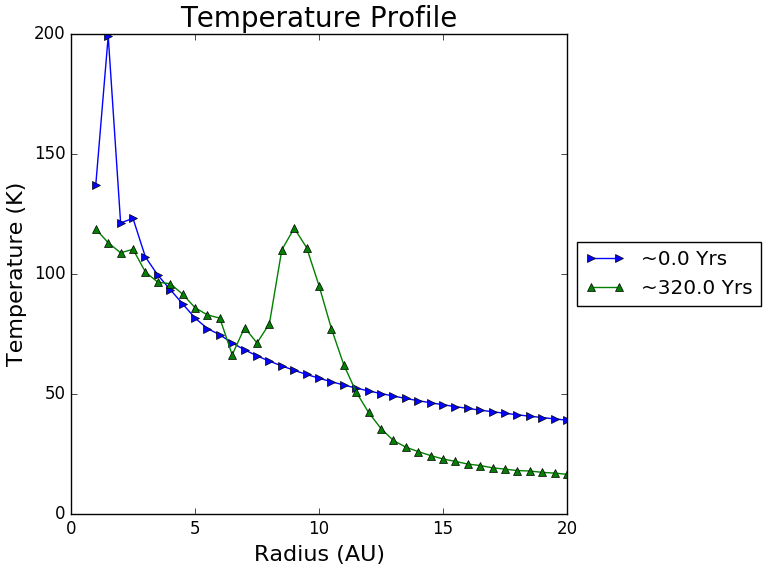}
  \caption{Temperature profile of the {\scshape starrad} static irradiated disk over \textasciitilde300 years of evolution, corresponding to \textasciitilde30 disk orbits at 10AU.}
  \label{fig-starrad-temp}
  \end{center}  
 \end{figure}

\begin{figure}[h!]
  \begin{center}  
\includegraphics[width=.97\columnwidth]{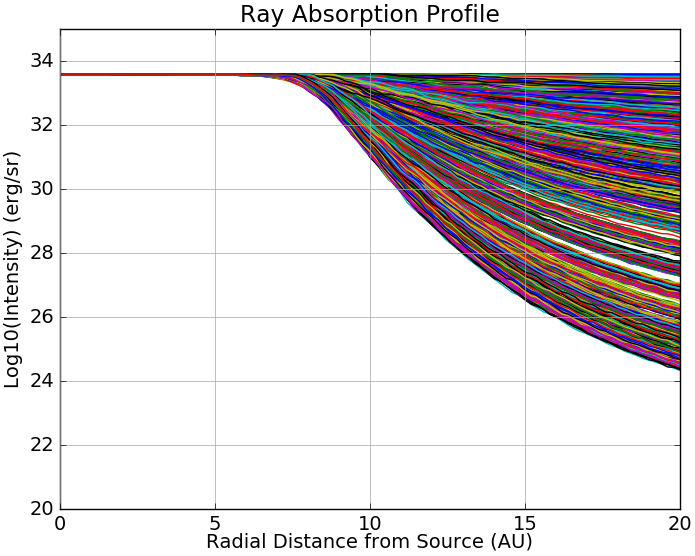}
  \caption{
  Absorption profiles for rays. The initial low density below 6AU provides negligible absorption, while above 6AU the energy (per unit solid angle) of each ray steadily declines as it is absorbed by the denser regions of the disk.
  }
  \label{fig-starrad-rays}
  \end{center}  
 \end{figure}

\begin{figure}[h!]
\begin{center}  
\includegraphics[width=.97\columnwidth]{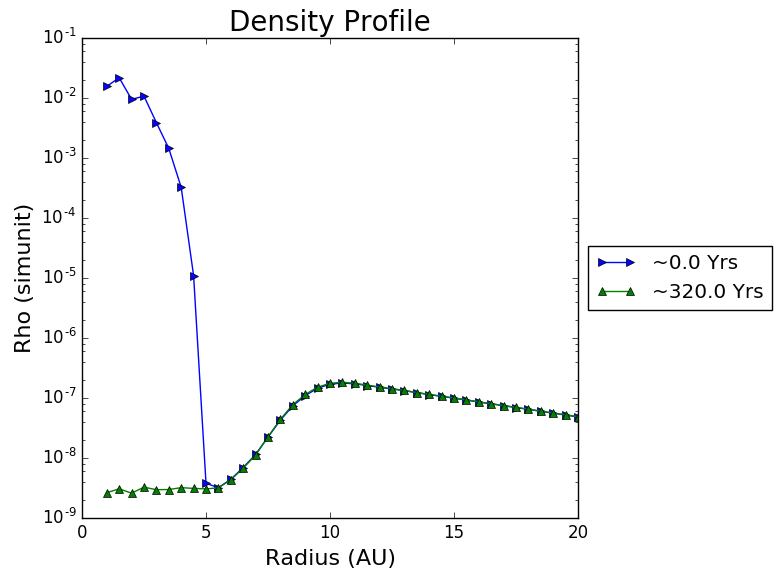}
  \caption{Density profile of the {\scshape starrad} initial conditions and irradiated disk after \textasciitilde300 years of evolution. This does not change significantly after an initial readjustment phase in the sparse inner region of the initial disk.}
  \label{fig-starrad-rho}
  \end{center}  
 \end{figure}

Fig.~\ref{fig-starrad-profiles} (right) shows the specific irradiation of the disk as calculated by {\scshape starrad}, i.e. the energy absorbed per unit time per unit mass, along with a temperature map after heating - the dense inner disk clearly shielding the outer disk in the midplane. The temperature profile in Fig.\ref{fig-starrad-temp} demonstrates that the {\scshape starrad} module combined with cooling quickly reaches a steady-state to reasonable values for protoplanetary disks. The incline in temperature in the 6-10AU range results from high absorption reflecting the optical depth profile of the initial conditions. A plot of ray absorption in Fig. \ref{fig-starrad-rays} confirms negligible absorption within a \textasciitilde6 AU radius, matching the density profile seen in Fig. \ref{fig-starrad-rho}. Beyond the 9 AU radius, fast cooling results in the subsequent temperature decline, this is expected to be more moderate when coupled with FLD.  

While the current tests utilize a static disk, to simplify validation of the radiative transport, {\scshape starrad} is also able to run with hydrodynamics and will be used in upcoming production runs; indeed in the current run the hydrodynamics are active (recomputing density/pressure at each step) however accelerations are turned off, hence a static disk.  Note that the initial adjustment of the disk from time t$=$0 is due to initialization procedure (described above) applied to enforce hydrostatic equilibrium to the disk, whose inner region consists of few particles due to a hole by construction (Fig 
\ref{fig-starrad-rho}, $<5$AU).  We have verified the accuracy of {\scshape starrad} in terms of energy conservation and heating rates; more rigorous verification of its accuracy is still underway. 
Full details of the {\scshape starrad} module will be the subject of a forthcoming dedicated paper [in preparation].

\section{Conclusions and future work}
\label{conclusions}

We have presented the first results of coupling the {\scshape diaphane} library to different hydrodynamics codes for simulating diffusion through a 3D-periodic uniform distribution of gas. The very good agreement among all codes is encouraging and proves that the porting of the library is successful, independently of the underlying hydrocode. Additional modules like {\scshape starrad} (a radiative transfer solver based on ray casting) and {\scshape asl} (a spectral leakage treatment for neutrino radiation) are currently being implemented and tested.

The {\scshape diaphane} library allows multiple radiation and neutrino transport algorithms to be used by a generic hydrodynamic code by calling portable modules using a common API.  This functionality has the potential for a transformative improvement in the modelling of complex astrophysical processes on scales spanning a vast range from kilometer scale black holes to star and galaxies and beyond.

The current implementation that calls the library modules using a fluid element basis will be improved to allow optimization by vectorization and accelerators such as GPUs or MIC processors.  Such work is necessary to use the library on future Exascale machines.  Having a GPU capable code, \small{ChaNGa}, already supported by our collaboration will be conducive to future GPU porting work.  \small{ChaNGa}\footnote{http://www-hpcc.astro.washington.edu/tools/changa.html} is the GASOLINE hydro code, which is part of the library, re-engineered to work with a \small{CHARM++} parallel environment for better parallel performance owing to dynamic load balancing.
Future versions of the library are planned to be capable of computing on a spatially local domain basis with a single call rather than just on one fluid element at a time.  This would allow an accelerator to compute the RT/NT processes on its own set of particles independently of the main simulation code.  Further optimization would be possible if the library handles operations like neighbor finding internally.  The strategy for this development will be to extend our use of  the Machine Dependent Layer (MDL), the parallel library already used by {\scshape starrad}.  MDL has recently been updated to be efficient at hybrid parallel computing MPI and pthreads on node and GPU and was used for a two trillion particle cosmological gravity simulation \cite{Potter2016}.  Using the MDL is meant to simplify the work of porting the library to a new code (and onto new HPC hardware) by abstracting the parallel communication out of the library physics routines and into the interface layer. 
MDL is not limited to particle-based codes; a general MDL
that interfaces also with grid-based codes such as {\small RAMSES} is at an advanced state of development, as part of an effort lead by Joachim Stadel.

We plan to extend our simulation support to grid-based codes, beginning with {\small RAMSES}.  The development effort will be mainly focused on the common API layer and, even more, in its auxiliary wrapper functions.  The prior development of an MDL version of {\scshape ramses} will help with this effort.  In some simple algorithms such as FLD, we plan to provide both a grid-based and a particle-based module, depending on which type of code is running. However, some particle-based methods may not translate easily to grid-based methods.  We also will allow the option of converting grid cells into pseudo-particles, so that a particle-based method can be applied.

A future goal is to incorporate adaptive coupling of the multiple modules.  One method that we are exploring for this coupling is the Fuzzy Domain Method (FDM) \cite{GanderMichaud2014}.  This is a way of computing an approximate solution in intermediate regimes where no algorithm is ideal and can provide a better solution than simple interpolation.  Conceptually, FDM is well suited to couple solutions formally analogous to those in moment-based methods, namely for which there is no underlying discretization of the radiation field using rays or domains but rather the radiation field is treated as continuum.  Hence one can imagine providing two limiting solutions in the optically thin and optically thick regime, suited for certain physical problems, and use FDM to interpolate.  
FDM has already been utilized for neutrino transport in core-collapse supernova simulations with regard to neutrinos that are trapped (i.e. high optical depth) or stream outward (i.e. low optical depth) from the explosion \cite{Berninger2014}.  
Often intermediate regimes will have no accurate technique available.  This is where the Fuzzy Domain Techniques hold promise beyond what a simple interpolation of two algorithms would yield.  Although it is still under development, FDM will be an important feature of the library.  We note that FDM is a novel approach that will enable multiple RT algorithms to operate within a running simulation in an automated and mathematically motivated way.  The switching and coupling between different modules will be handled internally by the library, using for example, the local optical depth for the switching or coupling.

Future work, which will go in the second release of the library, will see the implementation of an M1 moment-based method (originating from the existing implementation in RAMSES; \cite{Rosdahl2013}) and the implementation of TRAPHIC.  With these and the existing modules, we will cover all the main categories of RT approximations and schemes available in astrophysics.  This means that when another team will want to implement a new scheme this will inevitably have commonalities with one of those already implemented, enabling to exploit the already existing API and auxiliary functions and those requiring little added work.  Hence our strategy is by construction aimed at maximizing the output to the community at large. 

Development of the neutrino transport modules is still ongoing. Nevertheless, the {\scshape asl} scheme, as well as the IDSA method, have been already ported to different SPH codes and mesh codes, currently both in production stages.  This knowledge is transferable to their inclusion in {\scshape diaphane} and we are benefiting considerably from it.

\section{Availability}
   The current version of the Diaphane library is in early use exclusively within the collaboration.  A first public version is expected to be released as an open-source code in 2018, to be submitted for publication to {\it Computer Physics Communications} as a {\it Computer Programs in Physics} paper.

\section{Acknowledgments}
This work has been supported by the Swiss Platform for Advanced Scientific Computing (PASC, http://www.pasc-ch.org/) project DIAPHANE (DR and RC) and by the European Research Council (FP7) under ERC Advanced Grant Agreement No. 321263 - FISH (RC). 
We would also like to thank our collaborators Mattias Liebend\"orfer, Martin Gander, Doug Potter, Alireza  Rahmati, Joachim Stadel, Romain Teyssier, Gabriele Ciaramella, and James Wadsley, for many insightful discussions.
The authors acknowledge the support of sciCORE (http://scicore.unibas.ch/) scientific computing core facility at University of Basel, the Swiss National Supercomputing Center CSCS (http://cscs.ch), and S$^{3}$IT at the University of Z\"{u}rich, where these calculations were performed.





\bibliographystyle{cpc}
\bibliography{sigproc}


%
%




\end{document}